\documentclass[pra,english,reprint,nofootinbib]{revtex4-1}
\usepackage[T1]{fontenc}
\usepackage[utf8]{inputenc}
\usepackage{textcomp}
\usepackage{mathrsfs}
\usepackage{mathtools}
\usepackage{amsmath}
\usepackage{amssymb}
\usepackage{graphicx}
\usepackage{babel}
\usepackage{color}
\usepackage{siunitx}
\usepackage[normalem]{ulem}
\begin{document}
\addtolength{\textheight}{2mm}
\global\long\def\ket#1{\left|#1\right\rangle }%
\global\long\def\bra#1{\left\langle #1\right|}%
\global\long\def\bk#1#2{\left\langle #1\right|\left.#2\right\rangle }%
\global\long\def\bok#1#2#3{\left\langle #1\right|#2\left|#3\right\rangle }%
\global\long\def\gk#1{\left\{  #1\right\}  }%
\global\long\def\k#1{\left(#1\right)}%
\global\long\def\ek#1{\left[#1\right]}%
\global\long\def\abs#1{\left|#1\right|}%
\global\long\def\dk#1{\left\langle #1\right\rangle }%
\global\long\def\Re{\text{Re}\,}%
\global\long\def\Im{\text{Im\,}}%
\global\long\def\sumprime{\sideset{}{^\prime}\sum}

\title{Real space dynamics of attractive and repulsive polarons in Bose-Einstein
condensates}
\author{Moritz Drescher}
\author{Manfred Salmhofer}
\author{Tilman Enss}
\affiliation{Institut für Theoretische Physik, Universität Heidelberg, D-69120
Heidelberg, Germany}
\begin{abstract}
  We investigate the formation of a Bose polaron when a single
  impurity in a Bose-Einstein condensate is quenched from a
  non-interacting to an attractively interacting state in the vicinity
  of a Feshbach resonance. We use a beyond-Fröhlich Hamiltonian to describe
  both sides of the resonance and {a coherent-state} 
  variational ansatz to compute the time evolution of boson density
  profiles in position space. We find that on the repulsive side of
  the Feshbach resonance, {the Bose polaron performs long-lived oscillations, which is
  surprising given that the two-body problem has only one bound state coupled to a continuum. They
  arise due to interference between multiply occupied bound states and therefore can be only found with many-body
  approaches such as the coherent-state ansatz. This is a distinguishing feature of the Bose polaron compared to the Fermi polaron where
  the bound state can be occupied only once. We derive an implicit equation
  for the frequency of these oscillations and show that it can be approximated by the energy of the
  two-body bound state. Finally, we consider an impurity introduced at non-zero velocity and find
  that, on the repulsive side,} it is periodically slowed down or
  even arrested before speeding up again.
\end{abstract}
\maketitle

\section{Introduction}

The polaron is a general concept of many-body physics that naturally
arises in different fields like solid state physics and the theory
of ultracold gases. While it has long been used to describe electrons
in a crystal lattice, only recent experimental advances allowed one
to realize polarons in ultracold gases. Here, the Feshbach resonance
allows for a high level of control and in particular for realizing
the strong-coupling regime, which could not be done before. This gives
access to interesting phenomena such as self-localization and bubble
formation. Moreover, the interaction can be changed abruptly, which
allows for the investigation of the dynamics. Combined with the possibility
of direct imaging, this allows to view polaron formation in position
space, which is crucial for the physical intuition and interpretation
of the time evolution.

The concept of polarons was originally invented by Landau \cite{Landau1933}.
He showed that an electron in a crystal lattice interacts with the
surrounding atoms in such a way that it can be described as a quasiparticle
with a higher effective mass, moving through free space. Describing
the lattice deformations induced by the electron as phonons, the polaron
can be imagined as an electron carrying a cloud of phonons around
it. A very similar picture arises in ultracold bosonic gases: According
to Bogoliubov theory, the elementary excitations of a BEC are phonons
as well, so when an impurity is moving through the gas, the situation
is analogous to that of an electron in a crystal. But in an ultracold
gas, it is possible to tune the interaction between the particles
via a Feshbach resonance and in particular to investigate the regime
of strong coupling between impurity and host bosons.

A number of different theoretical approaches has been used to
investigate different aspects of the Bose polaron. In 1954, Fröhlich
introduced a Hamiltonian which is commonly used to study polarons
\cite{Frohlich1954}.  It can be recovered from Bogoliubov theory with
one further approximation \cite{Girardeau1961}. This was first done
for ultracold gases in \cite{Tempere2009}, where the ground state
properties were studied using a variational ansatz due to Feynman
\cite{Feynman1955}. This ansatz works well for all couplings in the
original case of electrons in a lattice, but in the ultracold gas, the
regularization of the contact interaction leads to errors when the
coupling becomes strong. This was discussed in \cite{Vlietinck2015}
with Diagrammatic Monte Carlo calculations.  These give access to the
ground-state properties and are computationally intensive but
numerically exact and provide valuable benchmarks for other methods. A
coherent-state variational ansatz originally due to Lee, Low and Pines
(LLP) \cite{Lee1953} has been used to study dynamical properties
\cite{Shashi2014, Shchadilova2016}. It neglects entanglement in
momentum space and is considered best for heavy impurities and weak
couplings. More quantum fluctuations have been taken into account by a
renormalization group technique for the ground state \cite{Grusdt2015,
  Grusdt2016all} and the dynamics \cite{Grusdt2018} and by the
correlated gaussian wave function ansatz \cite{Shchadilova2016a}, as
well as a Hartree-Fock-Bogoliubov description \cite{Kain2016}.  There
are some more works related to the Fröhlich Hamiltonian
\cite{Casteels2010,Casteels2012,Nielsen2018}; for a review, see
\cite{Grusdt2015a}. The Bose polaron exhibits characteristic
signatures also at finite temperature \cite{Levinsen2017,
  Guenther2018}.

The interaction term in the Fröhlich Hamiltonian is, however, just an
approximation in the case of ultracold gases and higher order terms
become important in the regime of strong coupling. This was first
observed in \cite{Rath2013}, where a T-matrix approximation was used 
{(for a real-time version, see \cite{Volosniev2015}).
Subsequently, a number of approaches have been applied to the fully
interacting model within Bogoliubov theory \cite{Shchadilova2016,
Grusdt2017, Li2014, Christensen2015, Schmidt2018}.}
In the one-dimensional case, some analytical results for heavy impurities have been obtained
\cite{Kain2018} and phonon-phonon interactions beyond Bogoliubov
theory have been considered \cite{Grusdt2017a}.

Approaches not based on Bogoliubov theory are more limited in number:
Quantum Monte Carlo calculations \cite{Ardila2015} provide exact
ground states for a limited number of parameters. Coupled
Gross-Pitaevskii equations \cite{Astrakharchik2004, Bruderer2008,
  Blinova2013} can describe the spatial deformation of the BEC and the
phenomena of self-localization and the bubble polaron but work on a
mean-field level. A variational approach which treats the molecular
state as an independent quasi-particle has been used to investigate
three-body bound states \cite{Levinsen2015,Christensen2015}.
{In one dimension, the Bose polaron problem can be solved
exactly in certain limiting cases \cite{McGuire1965} and the general case has been addressed
by related techniques \cite{Volosniev2017}, but these methods do not
carry over to three dimensions.}

Experimentally, Bose polarons in ultracold gases have been observed
with a focus on absorption spectra and decoherence
\cite{Catani2012, Scelle2013, Hu2016, Jorgensen2016, Camargo2018},
for which some theoretical predictions have been made. Direct imaging
experiments on the other hand are still in preparation and there have
been few theoretical results concerning the real space dynamics of
the bose polaron: the Monte Carlo calculations in \cite{Ardila2015}
include density profiles but only statically for ground states while
the Gross-Pitaevskii method in \cite{Blinova2013} considered a repulsive
interaction. This is different from an attractive interaction with
a positive scattering length in that it does not feature a bound state.

In this paper, we investigate the dynamics of polaron formation when
an initially non-interacting impurity is quenched to an attractively
interacting state. This situation has been studied before to compute
radio-frequency absorption spectra \cite{Shashi2014,Shchadilova2016}
and, on the attractive side of the Feshbach resonance, polaron
trajectories \cite{Grusdt2018}, as well as pre-thermalization dynamics
\cite{Lausch2018}. Here, we focus on two new aspects: We compute the
density profile of the BEC around the impurity as a function of time
and thus view the formation of the polaron in position space.  This
can be directly measured with current imaging technologies, and
corresponding experiments are in preparation. On the other hand, we
investigate the repulsive side of the Feshbach resonance where the
scattering length is positive. Here, a two-body bound state exists and
its interplay with the polaron leads to new effects, in particular
characteristic oscillations and a depletion of the boson density in a
halo around the impurity. These were inaccessible to many previous
works based on the Fröhlich Hamiltonian, which depends only on the
modulus of the scattering length and cannot describe the bound states.
Our study, instead, uses the extended Hamiltonian including
higher-order terms in the interaction{.
The dynamics are computed by applying a coherent-state ansatz. We find
oscillations on the repulsive side of the Feshbach resonance which arise as a result of multiply
bound states. This demonstrates the necessity to use a truly many-body ansatz such as the
coherent-state ansatz.}

The paper is organized as follows. In Sec.~\ref{sec:Model}, we review
the construction of the Hamiltonian and the variational ansatz starting
from Bogoliubov theory and discuss the stationary solution. Section
\ref{sec:Results} contains the results for the time evolution after
a quench: we start with the case of an impurity initially at rest
and compute boson density profiles as well as the total number of
bosons gathering around the impurity. {We then present an analytical study that demonstrates the
reason for the long-lived oscillations that occur on the repulsive side of the Feshbach resonance
and provide a way to compute their frequencies.}
Finally, we investigate the influence of a non-zero initial velocity and compute
polaron trajectories.

\section{Model\label{sec:Model}}

Our starting point is the Hamiltonian of a single impurity in a bath of
bosons
\begin{align*}
H= & \frac{\hat{p}_{I}^{2}}{2m_{I}}+\sum_{\boldsymbol{k}}\frac{k^{2}}{2m_{B}}a_{\boldsymbol{k}}^{\dagger}a_{\boldsymbol{k}}\\
 & +\frac{1}{2V}\sum_{\boldsymbol{k},\boldsymbol{q},\boldsymbol{p}}V_{BB}\k{\boldsymbol{p}}\,a_{\boldsymbol{k+p}}^{\dagger}a_{\boldsymbol{q-p}}^{\dagger}a_{\boldsymbol{k}}a_{\boldsymbol{q}}\\
 & +\int d^{3}\boldsymbol{x}\,V_{IB}\k{\boldsymbol{x}-\hat{\boldsymbol{x}}_{I}}n_{B}\k{\boldsymbol{x}}
\end{align*}
where $m_{I}$ and $m_{B}$ are the masses of impurity and bosons,
$\boldsymbol{\hat{p}_{I}}$ and $\boldsymbol{\hat{x}_{I}}$ the impurity
momentum and position operators and $a_{\boldsymbol{k}}^{\k{\dagger}}$
the bosonic creation and annihilation operators. $n_{B}(\boldsymbol{x})=a_{\boldsymbol{x}}^{\dagger}a_{\boldsymbol{x}}$
is the boson density and $V_{BB}$ and $V_{IB}$ are the boson-boson
and impurity-boson interaction potentials. Our derivation follows
Shchadilova et al.~\cite{Shchadilova2016}.

Since we are dealing with just one impurity, it is convenient to go
to relative coordinates. This is achieved by the {exact} canonical transformation
$\exp\k{iS}$ where
\[
S=\boldsymbol{\hat{x}_{I}}\cdot\sum_{\boldsymbol{k}}\boldsymbol{k}a_{\boldsymbol{k}}^{\dagger}a_{\boldsymbol{k}}\,.
\]
It is known as the Lee-Low-Pines (LLP) transformation
\cite{Lee1953}, see also \cite{Girardeau1961}. Its effect on the
operators is
\begin{align*}
e^{iS}\boldsymbol{\hat{p}_{I}}e^{-iS} & =\boldsymbol{\hat{p}}-\sum_{\boldsymbol{k}}\boldsymbol{k}a_{\boldsymbol{k}}^{\dagger}a_{\boldsymbol{k}}\,, & e^{iS}a_{\boldsymbol{k}}^{\dagger}e^{-iS} & =e^{i\boldsymbol{\hat{x}_{I}}\cdot\boldsymbol{k}}a_{\boldsymbol{k}}^{\dagger}\,,\\
e^{iS}a_{\boldsymbol{x}}^{\dagger}e^{-iS} & =a_{\boldsymbol{x}+\boldsymbol{\hat{x}_{I}}}^{\dagger}\,, & e^{iS}a_{\boldsymbol{k}}e^{-iS} & =e^{-i\boldsymbol{\hat{x}_{I}}\cdot\boldsymbol{k}}a_{\boldsymbol{k}}\,.
\end{align*}
Note that formally $\boldsymbol{\hat{p}}=\boldsymbol{\hat{p}_{I}}$,
but we have dropped the index after the transformation since the physical
meaning is not the impurity but the total momentum. It is, of course,
conserved and can be replaced by the initial impurity momentum $\boldsymbol{p_{0}}$
such that the transformed Hamiltonian reads
\begin{align*}
H_{\text{LLP}} & =\frac{\k{\boldsymbol{p_{0}}-\sum_{\boldsymbol{k}}\boldsymbol{k}a_{\boldsymbol{k}}^{\dagger}a_{\boldsymbol{k}}}^{2}}{2m_{I}}+\sum_{\boldsymbol{k}}\frac{k^{2}}{2m_{B}}a_{\boldsymbol{k}}^{\dagger}a_{\boldsymbol{k}}\\
 & +\frac{1}{2V}\sum_{\boldsymbol{k},\boldsymbol{q},\boldsymbol{p}}V_{BB}\k{\boldsymbol{p}}\,a_{\boldsymbol{k+p}}^{\dagger}a_{\boldsymbol{q-p}}^{\dagger}a_{\boldsymbol{q}}a_{\boldsymbol{k}}\\
 & +\int d^{3}\boldsymbol{x}\,V_{IB}\k{\boldsymbol{x}}n_{B}\k{\boldsymbol{x}}\,.
\end{align*}
This transformation has simplified the interaction term and replaced
the impurity momentum by the difference of total momentum and boson
momentum. Here, fourth-order terms in the boson operators appear
unless the impurity is taken to be infinitely heavy, i.e., stationary.
A delocalized impurity thus induces effective interactions between
the bosons.

\subsection*{Bogoliubov Theory}

We use Bogoliubov theory which pre-supposes Bose-Einstein-Condensation
in the $\boldsymbol{k}=0$ mode and approximates the low-temperature
behaviour by discarding terms in 3rd and 4th order of boson operators
with $\boldsymbol{k}\ne0$. The resulting bosonic part of the Hamiltonian
is diagonalized by the Bogoliubov transformation $b_{\boldsymbol{k}}^{\dagger}=\cosh\k{\varphi_{k}}a_{\boldsymbol{k}}^{\dagger}-\sinh\k{\varphi_{k}}a_{-\boldsymbol{k}}$
with $\exp\k{4\varphi_{k}}=\xi^{2}k^{2}/(2+\xi\text{\texttwosuperior}k\text{\texttwosuperior})$
and the healing length
\[
\xi=\frac{1}{\sqrt{8\pi a_{BB}n_{0}}}\,.
\]
Up to a constant energy offset,
\begin{align*}
H_{\text{Bog}}= & \frac{\k{\boldsymbol{p_{0}}-\sum_{\boldsymbol{k}}\boldsymbol{k}b_{\boldsymbol{k}}^{\dagger}b_{\boldsymbol{k}}}^{2}}{2m_{I}}+\sideset{}{^{\prime}}\sum_{\boldsymbol{k}}\omega_{k}b_{\boldsymbol{k}}^{\dagger}b_{\boldsymbol{k}}\\
 & +\int d^{3}\boldsymbol{x}\,V_{IB}\k{\boldsymbol{x}}n_{B}\k{\boldsymbol{x}}
\end{align*}
with phonon dispersion
\begin{align*}
\omega_{k} & =\frac{k}{2m_{B}\xi}\sqrt{2+\xi\text{\texttwosuperior}k\text{\texttwosuperior}}\,.
\end{align*}
 $a_{BB}$ and $a_{IB}$ are the scattering lengths of the potentials
$V_{BB}$ and $V_{IB}$. Finally, $n_{0}$ is the condensate density,
which is a free parameter in Bogoliubov theory. $\sum'$ means that
the sum runs over $\boldsymbol{k}\ne0$.

\subsection*{Contact interaction}

In a dilute ultracold gas, the range of interactions is small compared
to all other length scales. The effect of the interaction can
therefore be described by a single number, the scattering length,
while the precise shape of the potential does not matter and can be
chosen arbitrarily.  The most convenient choice is a zero-range
pseudopotential.  Taken literally, the Fourier transform of a delta
potential would correspond to a potential in momentum space that is
constant over an unbounded region, which does not make sense.
Instead, one constructs it as a scaling limit, first cutting off all
momentum sums at some large $\Lambda$, and then tuning the interaction
strength in the limit $\Lambda\to\infty$ such that the scattering
length remains fixed at the desired value.  The correctly regularized
interaction strength is then given by
\begin{equation}
\int d^{3}\boldsymbol{x}\,V_{IB}\k{\boldsymbol{x}}n_{B}\k{\boldsymbol{x}}=\frac{g_{IB}}{V}\sum_{\boldsymbol{k},\boldsymbol{q}}a_{\boldsymbol{k}}^{\dagger}a_{\boldsymbol{q}}\label{eq:Contact Interaction}
\end{equation}
\[
g_{IB}^{-1}=m_{\text{red}}\k{\frac{1}{2\pi a_{IB}}-\frac{2}{V}\sum_{\boldsymbol{k}}^{\Lambda}\frac{1}{k\text{\texttwosuperior}}}
\]
with $m_{\text{red}}$ being the reduced mass of impurity and bosons,
$m_{\text{red}}^{-1}=m_{I}^{-1}+m_{B}^{-1}$. Note that such a cutoff effectively
corresponds to an interaction with a non-zero range of order $1/\Lambda$.
The cutoff $\Lambda$ will be used implicitly in all sums and integrals throughout
the paper. Also note that instead
of a ``hard'' cutoff, one can also multiply the integrands by a
decaying function such as $\exp\k{-2k{{}^2}/\Lambda{{}^2}}$. This
leads to smoother results when the cutoff is not large enough for
perfectly converged behaviour.

In eq. (\ref{eq:Contact Interaction}), we still have to express the
boson operators $a_{\boldsymbol{k}}$ by phonon operators $b_{\boldsymbol{k}}$.
The result is
\begin{align}
\sum_{\boldsymbol{k},\boldsymbol{q}}a_{\boldsymbol{k}}^{\dagger}a_{\boldsymbol{q}}= & N_{0}+\sqrt{N_{0}}\sideset{}{^{\prime}}\sum_{\boldsymbol{k}}W_{k}\k{b_{\boldsymbol{k}}^{\dagger}+b_{\boldsymbol{k}}}\nonumber \\
 & +\sideset{}{^{\prime}}\sum_{\boldsymbol{k},\boldsymbol{q}}\cosh\k{\varphi_{k}+\varphi_{q}}b_{\boldsymbol{k}}^{\dagger}b_{\boldsymbol{q}}\nonumber \\
 & \phantom{\sideset{}{^{\prime}}\sum_{\boldsymbol{k},\boldsymbol{q}}}+\sinh\k{\varphi_{k}+\varphi_{q}}\frac{b_{\boldsymbol{k}}^{\dagger}b_{\boldsymbol{q}}^{\dagger}+b_{\boldsymbol{k}}b_{\boldsymbol{q}}}{2}\label{eq:density_expanded}
\end{align}
where $W_{k}=\exp\k{\varphi_{k}}$. Here, $N_{0}=n_{0}V$ is the number
of condensed bosons and we have approximated $\bok 0{\hat{N}}0\approx N_{0}$,
i.e. neglected the ground state depletion, which gives a constant
density shift \cite{PitaevskiiStringari} $\frac{1}{V}\bok 0{\hat{N}}0-n_{0}=\frac{\sqrt{2}}{12\pi^{2}}\xi^{-3}\approx0.01\xi^{-3}$.

Inserting (\ref{eq:density_expanded}) into $H_{\text{Bog}}$, we
obtain the final Hamiltonian

\begin{align}
H= & \frac{\k{\boldsymbol{p_{0}}-\sum_{\boldsymbol{k}}'\boldsymbol{k}b_{\boldsymbol{k}}^{\dagger}b_{\boldsymbol{k}}}^{2}}{2m_{I}}+\sideset{}{^{\prime}}\sum_{\boldsymbol{k}}\omega_{k}b_{\boldsymbol{k}}^{\dagger}b_{\boldsymbol{k}}+g_{IB}n_{0}\nonumber \\
 & +g_{IB}\sqrt{\frac{n_{0}}{V}}\sideset{}{^{\prime}}\sum_{\boldsymbol{k}}W_{k}\k{b_{\boldsymbol{k}}^{\dagger}+b_{\boldsymbol{k}}}\nonumber \\
 & +\frac{g_{IB}}{V}\sideset{}{^{\prime}}\sum_{\boldsymbol{k},\boldsymbol{q}}\cosh\k{\varphi_{k}+\varphi_{q}}b_{\boldsymbol{k}}^{\dagger}b_{\boldsymbol{q}}\nonumber \\
 & \hphantom{\frac{g_{IB}}{V}\sideset{}{^{\prime}}\sum_{\boldsymbol{k},\boldsymbol{q}}}+\sinh\k{\varphi_{k}+\varphi_{q}}\frac{b_{\boldsymbol{k}}^{\dagger}b_{\boldsymbol{q}}^{\dagger}+b_{\boldsymbol{k}}b_{\boldsymbol{q}}}{2}\,.\label{eq:Final Hamiltonian}
\end{align}

The first two lines of (\ref{eq:Final Hamiltonian}) correspond to
a Fröhlich Hamiltonian, which is often used to study polarons. It
has also been used for polarons in ultracold gases, even though for
such systems the quadratic terms in the last two lines of eq. (\ref{eq:Final Hamiltonian})
are present. The so obtained results are still valid as long as the
coupling between impurity and host atoms is sufficiently weak. One
needs to take care however, that when using the Fröhlich Hamiltonian,
the regularized contact interaction $g_{IB}$ may not be used and
needs to be replaced by the result from the Born approximation $g_{IB}^{Fr}=2\pi a_{IB}/m_{\text{red}}$.
Near the Feshbach resonance, the quadratic terms become important
as pointed out in \cite{Rath2013,Shchadilova2016,Li2014,Christensen2015}.

\subsection*{Coherent state ansatz}

Also by Lee, Low and Pines \cite{Lee1953}, a variational ansatz for the Fröhlich
model was suggested, which approximates the ground state by a coherent
state. In the limit of infinitely heavy impurities, this ansatz becomes
exact. In \cite{Shashi2014}, a time dependent version of this ansatz
has been applied to the Bose polaron described by a Fröhlich Hamiltonian.
Subsequently, the same time-dependent ansatz has been applied to the
full Hamiltonian (\ref{eq:Final Hamiltonian}) in \cite{Shchadilova2016}.

Specifically, one considers wave functions of the form

\[
\ket{\alpha(t)}=\exp\k{\frac{1}{\sqrt{V}}\sideset{}{'}\sum_{\boldsymbol{k}}\alpha_{k}(t)b_{\boldsymbol{k}}^{\dagger}-h.c.}\ket 0
\]
and projects the Schrödinger equation onto the submanifold spanned
by these functions. Equivalent to this projection is the stationarity
of the functional
\[
\int dt\,\mathcal{L}(\alpha(t),\dot{\alpha}(t)):=\int dt\,\bok{\alpha}{i\partial_{t}-H}{\alpha}\,.
\]
with respect to the $\alpha_{k}$. Setting up the Euler-Lagrange equations
\footnote{$\frac{\partial}{\partial\overline{z}}=\frac{1}{2}\k{\frac{\partial}{\partial\Re z}+i\frac{\partial}{\partial\Im z}}$
denotes a Wirtinger derivative, which can be used instead of $\frac{\partial}{\partial\Re z}$
and $\frac{\partial}{\partial\Im z}$.} $\frac{\partial\mathcal{L}}{\partial\overline{\alpha}_{k}}-\frac{\partial}{\partial t}\frac{\partial\mathcal{L}}{\partial\dot{\overline{\alpha}}_{k}}=0$
results in the following differential equations, where we also took
the limit $V\rightarrow\infty$, replacing $\frac{1}{V}\sum'_{\boldsymbol{k}}$
by $\int\frac{d^{3}\boldsymbol{k}}{\k{2\pi}^{3}}$:
\begin{equation}
i\dot{\alpha}_{\boldsymbol{k}}=\k{\Omega_{k}-\frac{\boldsymbol{k}\cdot\boldsymbol{p_{I}}[\alpha]}{m_{I}}}\alpha_{\boldsymbol{k}}+W_{k}C_{1}[\alpha]+iW_{k}^{-1}C_{2}[\alpha]\label{eq:DiffEq}
\end{equation}
where
\begin{align*}
\Omega_{k} & =\frac{k^{2}}{2m_{I}}+\omega_{k}\\
\boldsymbol{p_{I}}[\alpha] & =\boldsymbol{p_{0}}-\int\frac{d^{3}\boldsymbol{k}}{\k{2\pi}^{3}}\,\boldsymbol{k}\abs{\alpha_{k}}^{2}\\
C_{1}[\alpha] & =g_{IB}\sqrt{n_{0}}+g_{IB}\int\frac{d^{3}\boldsymbol{k}}{\k{2\pi}^{3}}\,W_{k}\Re\alpha_{\boldsymbol{k}}\\
C_{2}[\alpha] & =g_{IB}\int\frac{d^{3}\boldsymbol{k}}{\k{2\pi}^{3}}\,W_{k}^{-1}\Im\alpha_{\boldsymbol{k}}\,.
\end{align*}
The initial value $\alpha_{\boldsymbol{k}}(0)=0$, i.e., $\ket{\alpha(0)}=\ket 0$,
corresponds to the situation of a quench from the phonon vacuum.

If the impurity is initially at rest, $\boldsymbol{p_{0}}=0$, then
$\boldsymbol{p_{I}}[\alpha]=0$ for all times due to spherical symmetry.
In this case, the equation becomes $\mathbb{R}$-linear and can be
written in the form
\begin{equation}
\begin{pmatrix}\Re\dot{\alpha}_{k}\vphantom{\alpha_{k}^{(s)}}\\
\Im\dot{\alpha}_{k}\vphantom{\alpha_{k}^{(s)}}
\end{pmatrix}=\begin{pmatrix}0 & \vphantom{\alpha_{k}^{(s)}}H^{(2)}\\
-H^{(1)} & \vphantom{\alpha_{k}^{(s)}}0
\end{pmatrix}\begin{pmatrix}\Re\big(\alpha_{k}-\alpha_{k}^{(s)}\big)\\
\Im\big(\alpha_{k}-\alpha_{k}^{(s)}\big)
\end{pmatrix}\label{eq:time evolution matrix}
\end{equation}
with a constant offset $\alpha^{(s)}$ (the stationary solution, see
below).

Our results are based on solving (\ref{eq:DiffEq}) numerically with
Verner's 8th-order Runge-Kutta scheme \cite{Verner2010}, using the
julia language \cite{Bezanson2014a} and the DifferentialEquations.jl
package \cite{Rackauckas2017}. In the $\boldsymbol{p_{0}}=0$ case,
we also diagonalize the matrix in (\ref{eq:time evolution matrix})
for comparison.
\begin{figure}
\includegraphics[width=0.8\columnwidth]{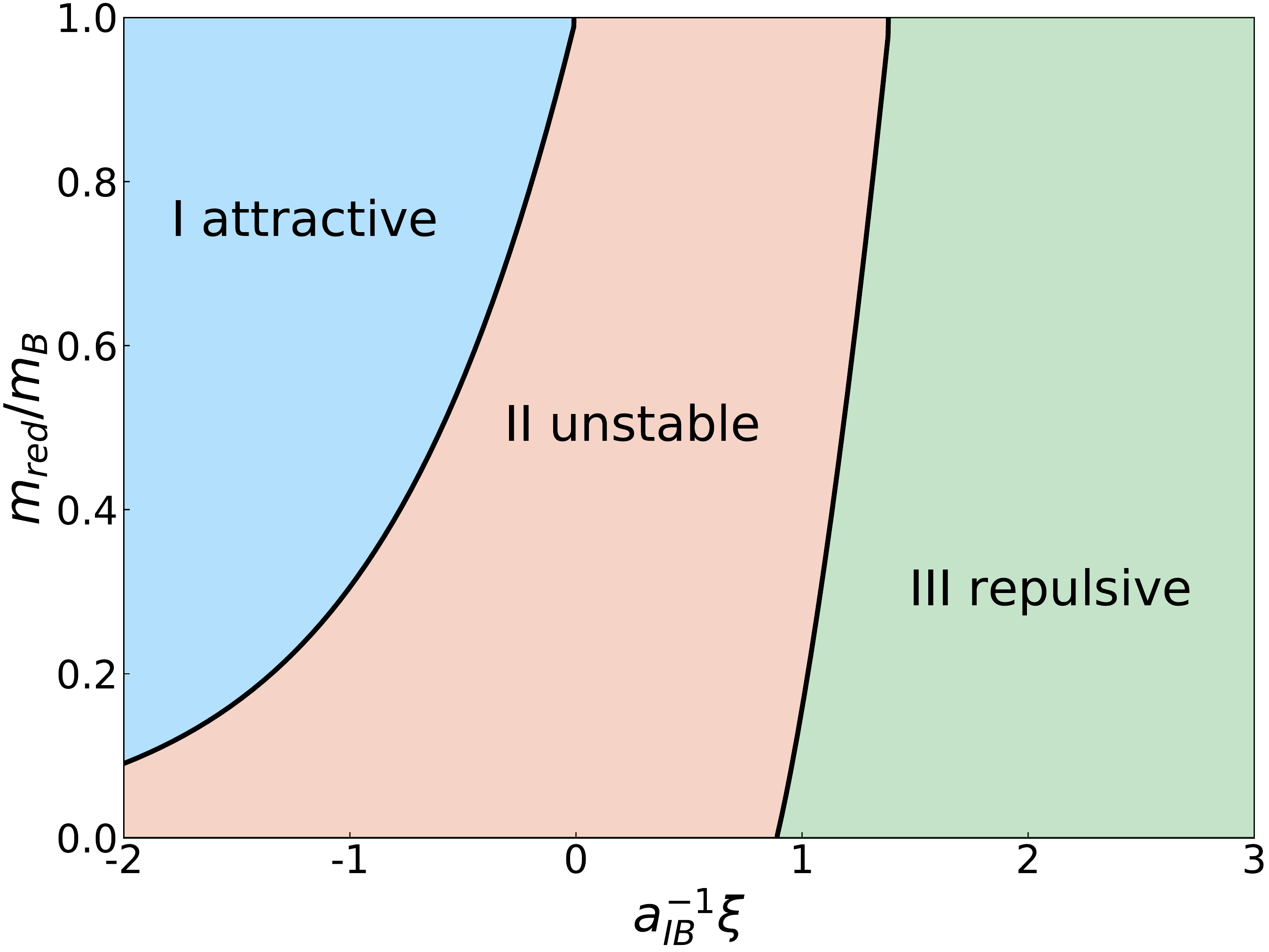}

\caption{\label{fig:Regimes}The three regimes of the coherent state ansatz,
exhibiting an attractive, unstable or repulsive polaron, separated
by the curves $a_{-}^{-1}$ and $a_{+}^{-1}$.}
\end{figure}

\subsection*{Stationary solution}

Before turning to the dynamical solutions, it is instructive to look
at the stationary solution obtained from $\dot{\alpha}_{\boldsymbol{k}}=0$
or equivalently $\frac{\partial\bok{\alpha}H{\alpha}}{\partial\alpha_{\boldsymbol{k}}}=0$.
One finds
\begin{equation}
\alpha_{\boldsymbol{k}}^{(s)}=-C_{1}\frac{W_{k}}{\Omega_{k}-\frac{\boldsymbol{k}\cdot\boldsymbol{p_{I}}}{m_{I}}}\label{eq:station=0000E4re L=0000F6sung}
\end{equation}
where $C_{1}$ and $\boldsymbol{p_{I}}$ are determined by the implicit
equations
\begin{align*}
C_{1} & =\sqrt{n_{0}}\k{g_{IB}^{-1}+\int\frac{d^{3}\boldsymbol{k}}{\k{2\pi}^{3}}\,\frac{W_{k}^{2}}{\Omega_{k}-\frac{\boldsymbol{k}\cdot\boldsymbol{p_{I}}}{m_{I}}}}^{-1}\\
\boldsymbol{p_{I}} & =\boldsymbol{p_{0}-}C_{1}^{2}\int\frac{d^{3}\boldsymbol{k}}{\k{2\pi}^{3}}\,\boldsymbol{k}\frac{W_{k}^{2}}{\k{\Omega_{k}-\frac{\boldsymbol{k}\cdot\boldsymbol{p_{I}}}{m_{I}}}^{2}}\,.
\end{align*}
Note that these quantities are UV convergent: For large $k$, one
has $W_{k}^{2}=1+\mathcal{O}(k^{-2})$ and $\Omega_{k}=\frac{k^{2}}{2m_{\text{red}}}+\mathcal{O}(1)$.
The {$C_1$-}integrand is thus $\frac{1}{\Omega_{k}}+\frac{\boldsymbol{k}\cdot\boldsymbol{p_{I}}}{\Omega_{k}^{2}m_{I}}+\mathcal{O}(k^{-4})$.
The first term cancels with the divergence of $g_{IB}^{-1}$, the
second vanishes by antisymmetry. The momentum integrand is $\frac{\boldsymbol{k}}{\Omega_{k}^{2}}+\mathcal{O}(k^{-4})$
and again, the first term is antisymmetric.

The integrals exist if and only if $p_{I}/m_{I}<c$, i.e., the stationary
impurity velocity must always be below the speed of sound $c=1/\sqrt{2m_{B}\xi}$.
For too large initial momenta, no stationary solution exists (the
same is true in the Fröhlich model where $C_{1}=2\pi a_{IB}\sqrt{n_{0}}/m_{\text{red}}$
is a constant, see \cite{Shashi2014}).

\begin{figure*}[!ht]
  \includegraphics[width=0.8\textwidth]{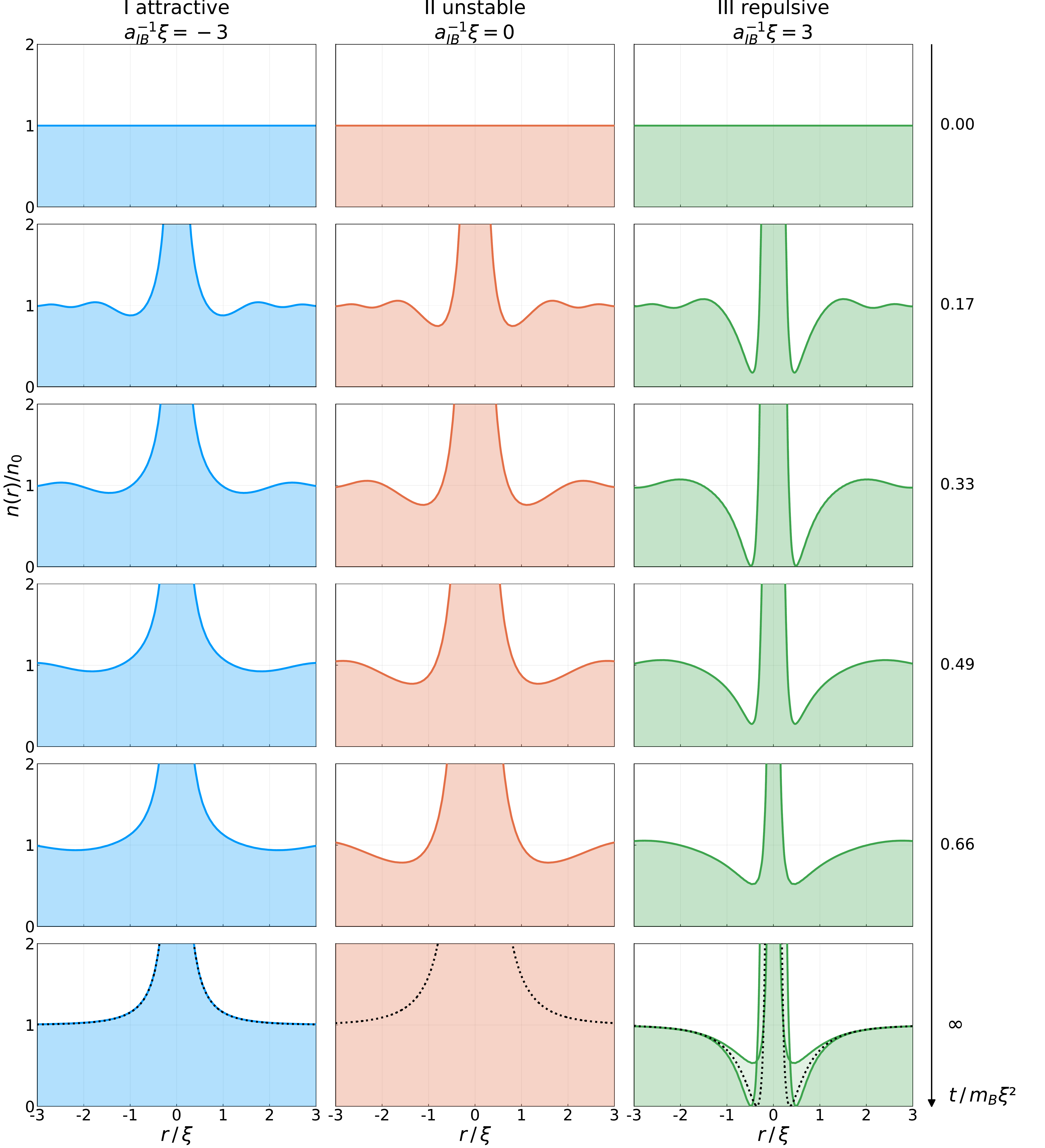}
  \caption{\label{fig:Density Profiles}Time series of density profiles
    of the bosonic bath. $r$ is the distance from the impurity. The
    densities are relative to the phonon vacuum state
    $\protect\ket 0$. The last row shows the long-time asymptotics as
    well as the stationary solution (\ref{eq:station=0000E4re
      L=0000F6sung}) (black dotted line). In the repulsive case, the
    system keeps oscillating between the two limit curves that are
    shown. Calculations were carried out at a gas parameter of
    $n_{0}a_{BB}^{3}=10^{-5}$, equal masses of bosons and impurity
    $m_{B}=m_{I}$ and with the impurity initially at rest,
    $\boldsymbol{p_{0}}=0$.  A soft momentum cutoff
    $e^{-2k{{}^2}/\Lambda{{}^2}}$ with $\Lambda=80\xi^{-1}$ was used.}
\end{figure*}
In the special case $\boldsymbol{p_{0}}=0$, one obtains $\boldsymbol{p_{I}}=0$,
\[
C_{1}=\frac{\sqrt{n_{0}}}{\frac{m_{\text{red}}}{2\pi}\k{a_{IB}^{-1}-a_{+}^{-1}}}
\]
and the stationary energy
\[
E^{(s)}=\frac{n_{0}}{\frac{m_{\text{red}}}{2\pi}\k{a_{IB}^{-1}-a_{+}^{-1}}}\,.
\]
Here, $a_{+}$ is one of two critical scattering lengths defined by
\[
a_{\pm}^{-1}=\frac{2\pi}{m_{\text{red}}}\int\frac{d^{3}\boldsymbol{k}}{\k{2\pi}^{3}}\,\k{\frac{2m_{\text{red}}}{k^{2}}-\frac{W_{k}^{\pm2}}{\Omega_{k}}}\,,
\]
which satisfy $a_{-}<0<a_{+}$. They were found in \cite{Grusdt2017}
and delimit three regions of different stability of the stationary
solution. This will be reflected in the convergence behaviour of observables
in our dynamical analysis:
\begin{enumerate}
\item $a_{IB}^{-1}<a_{-}^{-1}$: The stationary point is a minimum, observables
converge. This is the region where the attractive Bose polaron is
expected to form.
\item $a_{-}^{-1}<a_{IB}^{-1}<a_{+}^{-1}$: The stationary point behaves
like a saddle point, the system is dynamically unstable. This is well
understood as coming from phase fluctuations growing without bounds
\cite{Grusdt2017}.
    {This behaviour is unphysical and means that the approach cannot cover very strong
    couplings.
    }
\item $a_{+}^{-1}<a_{IB}^{-1}$: The stationary point behaves like a maximum,
observables are oscillating. In this region, the stationary solution
is usually interpreted as a repulsive polaron due to its positive
energy, while at negative energies, a molecular state is expected.
    {We will explain the reason for these oscillations and provide an estimate of the
    frequency.
    }
\end{enumerate}
In Fig.~\ref{fig:Regimes} we show how the boundaries of the three
regimes change with the reduced mass. For the case of light impurities,
$m_{\text{red}}\rightarrow0$, the unstable region grows. Here, quantum
fluctuations become especially important as the impurity is delocalized.

\subsection*{Applicability of the method}
Two approximations were involved in the derivation.

The Bogoliubov approximation neglects
third and fourth order terms in the Bose-Bose interaction of non-condensed modes. This is justified
if most of the particles are condensed since then, the coupling of excited to condensed modes
outweighs the coupling between different excited modes.

The coherent state ansatz, on the other hand, is a product state ansatz and as such, it neglects
correlations between different phonon modes. This is as well justified if the number of excited
particles is small. Note that the coherent state ansatz is closely related to Gross-Pitaevskii theory since
it corresponds to a replacement of a quantum field with a classical field and a coherent state in
momentum space is equivalent to one in real space.

In a weakly interacting Bose gas without an impurity, the condensate depletion is indeed very small.
If the impurity is added, there is, however, the unstable region in
which the theory predicts attraction of an unlimited number of bosons.
In reality, fourth order terms in the Bose-Bose interaction would prevent this.
Both in the attractive and repulsive regimes, however, the number of bosons attracted by the impurity will remain on the
order of only one to ten, as we show below, such that the theory is valid here.

{
For an estimate of the time scale on which the results can be trusted, observe that beyond Bogoliubov
theory, the decay time of phonons due to phonon-phonon interactions is proportional to the inverse
square root of the gas parameter:  $\tau \sim (n_0a_{BB}^3)^{-{1 \over 2}} {m_B \xi^2 \over \hbar} \approx
300 {m_B \xi^2 \over \hbar}$ for a typical gas parameter of $n_0a_{BB}^3 = 10^{-5}$, where
the prefactor depends on the number of excited modes. We find below that all interesting effects
occur for short times up to $10 {m_B \xi^2 \over \hbar}$. For these times the beyond-Bogoliubov
corrections are negligible even for local boson excitation numbers of order 10.
}

\section{Results\label{sec:Results}}

\subsection{Time evolution of density profiles}

Fourier transforming the numerical solution of (\ref{eq:DiffEq}) back
to position space, we can compute the boson density at distance $r$
from the impurity. More precisely, since the impurity is itself a
quantum particle, the quantity to consider is correlation function
\begin{align*}
n(\boldsymbol{x}) & =\dk{\hat{n}_{B}(\boldsymbol{\hat{x}_{I}}+\boldsymbol{x})}\\
 & =\left<\hat{n}_{B}(\boldsymbol{x})\hat{n}_{I}(\boldsymbol{x})\right> & \text{(original frame)}\\
n(\boldsymbol{x}) & =\left<\hat{n}_{B}(\boldsymbol{x})\right>\,. & \text{(LLP frame)}
\end{align*}

Expressing the boson density by phonon operators and applying the
variational ansatz, $n(\boldsymbol{x})$ takes the following form
in terms of the coefficients $\alpha_{\boldsymbol{k}}$:
\begin{align}
n(\boldsymbol{x})= & \k{\sqrt{n_{0}}+\Re\mathscr{F}^{-1}\left(\alpha W\right)(\boldsymbol{x})}^{2}\nonumber \\
 & +\k{\Im\mathscr{F}^{-1}\left(\alpha W^{-1}\right)(\boldsymbol{x})}^{2}\label{eq:density}
\end{align}
where $\mathscr{F}^{-1}f(\boldsymbol{x})=\int\frac{d{{}^3}\boldsymbol{k}}{(2\pi){{}^3}}e^{i\boldsymbol{k}\cdot\boldsymbol{x}}f(\boldsymbol{k})$
denotes the transformation to position space.

Figure~\ref{fig:Density Profiles} shows the results for the three different
regimes:
\begin{enumerate}
\item Attractive regime: Bosons are gathering around the impurity and the
profile quickly converges to form the attractive Bose polaron. The
final shape matches precisely that of the stationary solution.
\item Unstable regime: The impurity keeps pulling in more and more bosons.
As mentioned before, this unphysical behaviour reflects the failure
of the Bogoliubov approximation when interactions are too strong.
Including phonon interactions, i.e., higher-order terms in the bosonic
operators $a_{\boldsymbol{k}}$, might prevent this.
\item Repulsive regime: Close to the impurity, the boson density is strongly
increased but there is a halo of reduced density around it. There
is no convergence to a ground state profile, but instead, the solution
keeps oscillating between two states of the coupled system of impurity
and surrounding condensate:
At some
times, the bath is completely depleted at a certain distance while
at other times, there is still about half the original density left.
\end{enumerate}
\begin{figure}
  \centering\includegraphics[width=1\columnwidth]{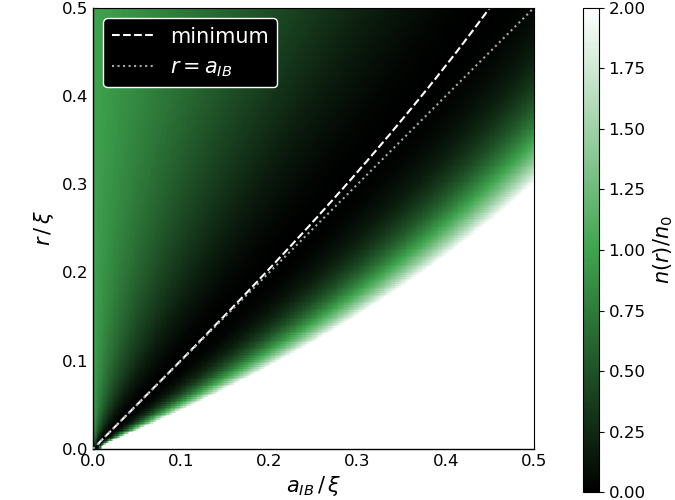}
  \caption{\label{fig:Density Heatmap}Stationary density profiles for
    $a_{IB}^{-1}>a_{+}^{-1}$.  The point of minimum density is at the
    distance of the impurity-boson scattering length
    $a_{IB}$. Parameters are as in Fig.~\ref{fig:Density Profiles}
    except that $\Lambda=600\xi^{-1}$ (to ensure
    $a_{IB}\ll\Lambda^{-1}$ even for weak coupling).}
\end{figure}
Comparing these results with the quantum Monte Carlo calculations
of the ground state profile in \cite{Ardila2015}, the results are
qualitatively similar, even though quantitatively slightly different
(our parameters correspond to $a_{IB}/a_{BB}=-20.94$, $\infty$ and
$20.94$).

\begin{figure*}
  \includegraphics[width=0.8\textwidth]{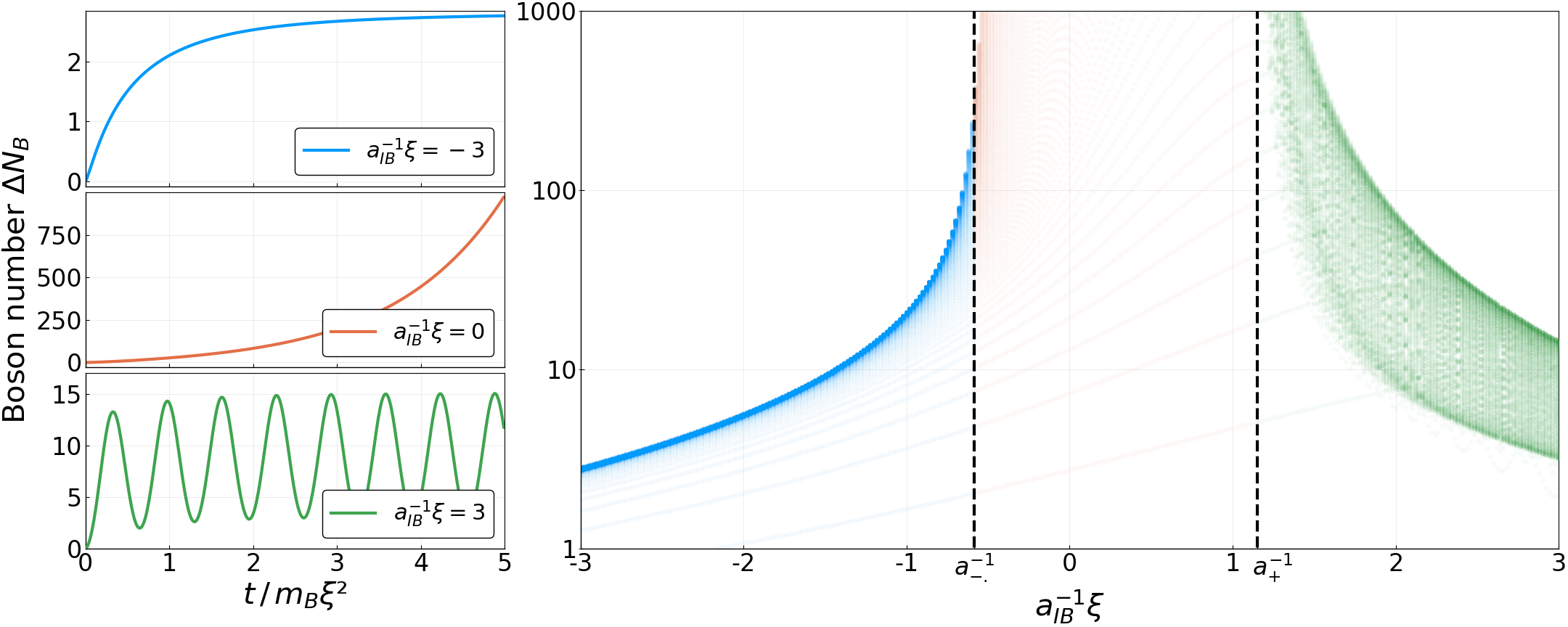}
  \caption{\label{fig:Boson Number Combined}The total number of bosons
    as function of scattering length and time. (a) For three different
    couplings.  (b) For a range of couplings. For each time point up
    to $t_{max}=60m_{B}\xi\text{\texttwosuperior}$, a low-opacity
    point is drawn. Where many points are on top of each other due to
    convergence or recurrence, this region becomes more opaque.
    Parameters are as in Fig.~\ref{fig:Density Profiles}, except
    that in (b) a hard momentum cutoff at $\Lambda=20$ was used.}
\end{figure*}
The complete depletion is a feature that is present even in the stationary
solution and for all scattering lengths above $a_{+}$: Since $\alpha^{(s)}$
is real, the last term in (\ref{eq:density}) vanishes and one can
always find an $r$ so that the first term vanishes as well. The length
scale on which the depletion takes place is given by the scattering
length, as shown in Fig.~\ref{fig:Density Heatmap}. This is not surprising
since this is the scale of the two-body bound state. The return to
the condensate density then happens on the order of the healing length
(not shown in the figure).

\subsection{Boson Number}

From the momentum space coefficients, we can compute the total change
in the number of bosons. This is not zero because the Bogoliubov theory
does not preserve particle number. It can be seen as a measure of
how many particles the impurity attracts in total. The formula is
\begin{align*}
\Delta N_{B}(t) & =\bok{\psi(t)}{\sum_{\boldsymbol{k}}n_{k}}{\psi(t)}-\bok 0{\sum_{\boldsymbol{k}}n_{k}}0\\
 & =\int\frac{d^{3}\boldsymbol{k}}{(2\pi)^{3}}\Big(\cosh(2\varphi_{k})\abs{\alpha_{\boldsymbol{k}}}^{2}+\\
 & \hphantom{=\int\frac{d^{3}\boldsymbol{k}}{(2\pi)^{3}}\Big(}\sinh(2\varphi_{k})\Re\alpha_{\boldsymbol{k}}\alpha_{-\boldsymbol{k}}\Big)\,.
\end{align*}
Results are shown in Fig.~\ref{fig:Boson Number Combined}a. The characteristics
of the three regimes - convergence, unbounded growth and oscillations
- are clearly visible. Note that in the repulsive case, the maxima
of the boson number correspond to the more extreme density profiles,
i.e., those with full depletion. Doing the same computations for many
different scattering lengths, we arrive at Fig.~\ref{fig:Boson Number Combined}b.
As the critical scattering lengths $a_{+}$ and $a_{-}$ are approached
from the attractive or repulsive regime, the total boson number grows
rapidly as well as the time to convergence. In the figure, this is
indicated by the fact that in these areas, the curves are still washed
out, therefore not yet converged.

{

\subsection{Discussion of the repulsive regime}

The presence of oscillations that do not decay is surprising to the physical
intuition, given that the two-particle problem features only one bound state and
a continuum of scattering states and one may wonder if this an artifact of one
of the approximations involved. In \cite{Li2014}, decaying oscillations were
predicted instead by applying a trial wave function of one impurity and at most
one phonon excitation. In \cite{Shchadilova2016}, this was contrasted with the same
approach that we use in this paper and which predicts stable oscillations. These 
were interpreted as ocurring between a many-body polaron branch and few-body bound
states. Here, we take a different point of view and claim that the repulsive polaron branch
plays no role. Instead, these oscillations occur between different multiply bound states.
This is a new feature of the Bose polaron in contrast to the Fermi polaron, where the bound
state can be occupied at most once.

{We demonstrate this by considering the simplified case of an infinitely heavy
impurity in a non-interacting BEC. But we emphasize that the latter restriction is not
necessary and undamped oscillations occur even in an exact solution of the full Bogoliubov-impurity
Hamiltonian. This is shown in appendix \ref{sec:OscillationMechanism}. Here, we restrict to the
non-interacting case, since the expressions are much simpler and the basic mechanism stands out clearer,
which is the same with and without Bose-Bose interactions.

For the situation considered here, 
}
the following Hamilonian is
exact (in the thermodynamic limit, to justify the substitution of $a_0$, $a_0^\dagger$
with $\sqrt{n_0}$):
\[
  H = \sideset{}{^{\prime}}\sum_k {k^2 \over 2m_B} a_k^\dagger a_k + g_{IB} \sqrt{n_0 \over V}
  \left( a_k^\dagger + a_k \right) + {g_{IB} \over V} \sideset{}{^{\prime}}\sum_{k,q} a_k^\dagger
  a_q
\]
The quadratic part can be easily diagonalized and yields the two-body spectrum. Even though
it has only one bound state, above Hamiltonian can lead to stable oscillations in observables.
This is surprising from the two-body point of view where a single eigenstate is coupled only to
the continuum, {such that oscillations dephase}. The difference lies in the linear terms in the Hamiltonian,
which lead to a shift of the creation and annihilation operators. Assume we have diagonalized
the quadratic part in terms of new operators $c_E = A_{Ek} a_k$, i.e. switched to the basis of two-body eigenstates:
\begin{align*}
  H &= \sum_E E c_E^\dagger c_E + E v_E c_E^\dagger + E \overline{v_E} c_E
\end{align*}
for some $A_{Ek}$ and $v_E$. The linear terms can be eliminated by the shift $d_E = c_E + v_E$:
\begin{align*}
  H &= \sum_E E d_E^\dagger d_E  + const.
\end{align*}
 This shift leads to a transformation of both the initial state
$\ket{0} = \ket{0}_c$ and observables, for instance the total particle number:
\begin{align*}
  \ket{0}_c &= \exp \left( \sum_E v_E d_E^\dagger - \overline{v_E} d_E \right) \ket{0}_d \\
  \hat N &= \sum_E c_E^\dagger c_E =  \sum_E (d_E^\dagger -\overline{v_E}) (d_E - v_E) \, .
\end{align*}
The time evolution of this operator is easily computed in the Heisenberg picture:
\begin{align*}
  \hat{N}(t) &= \sum_E \left( e^{iEt} d_E^\dagger - \overline{v_E} \right) \left( e^{-iEt} d_E -
  v_E \vphantom{d_E^\dagger} \right) \, .
\end{align*}

Now, if the quadratic part of the Hamiltonian has one bound state and a continuum of scattering
states---as is the case in the repulsive regime---the continuum part $E>0$ will dephase but oscillations
with the frequency of the bound state energy remain at long times:
\begin{align*}
  {}_c \bra{0} \hat{N}(t) \ket{0}_c &= \sum_E 2|v_E|^2 (1 - \cos(Et)) \\
     & \xrightarrow[]{t\rightarrow \infty} 2|v_{E_B}|^2 (1 - \cos(E_B t)) + \sum_{E>0} 2|v_E|^2
\end{align*}
where $E_B = -1/2m_{\text{red}}a_{IB}^2$.

{As we demonstrate in appendix \ref{sec:OscillationMechanism}, including
Bose-Bose interaction within Bogoliubov theory does not destroy this mechanism, since
the linear terms are still present} while the quadratic terms can be diagonalized by
means of a generalized Bogoliubov transformation.
On the other hand, third
and fourth order terms beyond Bogoliubov theory would likely lead to a damping of the oscillations.
But importantly, this damping rate is determined entirely by properties of the BEC while the
oscillation frequency is determined by the impurity-boson interaction. In experiments, these two 
time scales can be controlled independently and for weak Bose-Bose interactions, they will be
well distinguishable.

In this sense, we predict a damping reminiscent of the few-body
calculations \cite{Li2014}, but for a different reason and on different time scales:
In an ansatz with at most one phonon, the bound state couples only to the continuum and rapidly dephasing
oscillations are obtained. In an ansatz allowing an arbitrary number of excitations, coherent bound
states can be formed which decay only slowly because of Boson-Boson interactions.

This argument also shows that while our approach (and the Bogoliubov approximation in
general) is expected to be valid for all times on the attractive side of the Feshbach resonance,
it is valid on the repulsive side only as long as the damping has not set in, which is, however,
a large time scale for a weakly interacting Bose gas.
}

\subsection{Oscillation Frequencies}

In the case of an impurity initially at rest, the frequencies of the
oscillations in the repulsive regime can be predicted by making an
ansatz for the long-time solution. The coefficients $C_{1}$ and $C_{2}$
from the differential equation (\ref{eq:DiffEq}) show the same qualitative
behaviour as the other observables: convergence, divergence or oscillations,
according to the regime. We therefore make an asymptotic ansatz
\begin{align*}
C_{1} & =\sum_{\lambda}A_{\lambda}e^{\lambda t} \\
C_{2} & =\sum_{\lambda}B_{\lambda}e^{\lambda t}
\end{align*}
where the coefficients $\lambda$ can take finitely many complex values
with $\Re\lambda\ge0$. This covers all of the three cases: convergence
if only $\lambda=0$ is present, exponential growth if a $\lambda>0$
exists and oscillations for imaginary $\lambda$. The case $\Re\lambda<0$
would be interesting as well to describe the speed of convergence,
but the restriction to $\Re\lambda\ge0$ will prove necessary for
the calculation. Since $C_{1}$ and $C_{2}$ are real, we must have
$A_{\overline{\lambda}}=\overline{A_{\lambda}}$ and $B_{\overline{\lambda}}=\overline{B_{\lambda}}$
(the bar denotes complex conjugation).

Our aim is to derive conditions on $\lambda$ to be able to predict
the exponential growth rate or oscillation frequency of the physical
observables. We thus insert the ansatz into the differential equation
(\ref{eq:DiffEq}) with $\boldsymbol{p_{I}}[\alpha]=0$ and find the
solution

\[
\alpha_{k}(t)=s_{k}e^{-i\Omega_{k}t}+\sum_{\lambda}b_{k\lambda}e^{\lambda t}
\]

with the coefficients

\[
b_{k\lambda}=-\frac{W_{k}A_{\lambda}+iW_{k}^{-1}B_{\lambda}}{\Omega_{k}-i\lambda}
\]
and unknown $s_{k}$, which depend on the full history of the time
evolution.

{
This ansatz solves the projected Schrödinger equation asymptotically only if the values
of $\lambda$ are restricted to either $\lambda=0$ or the solutions of the implicit
equation
\begin{align}
  & \left(\Delta_{+}-\lambda^{2}\int\frac{d^{3}\boldsymbol{k}}{(2\pi)^{3}}\frac{W_{k}^{2}}{\Omega_{k}(\Omega_{k}^{2}+\lambda^{2})}\right)
 \nonumber \\
  \times &
  \left(\Delta_{-}-\lambda^{2}\int\frac{d^{3}\boldsymbol{k}}{(2\pi)^{3}}\frac{W_{k}^{-2}}{\Omega_{k}(\Omega_{k}^{2}+\lambda^{2})}\right)
  \nonumber \\
  = & -\lambda^{2}\left(\int\frac{d^{3}\boldsymbol{k}}{(2\pi)^{3}}\frac{1}{\Omega_{k}^{2}+\lambda^{2}}\right)^{2}\label{eq:lambda}
\end{align}
where we abbreviated
\[
  \Delta_\pm := {\mu \over 2\pi} \left(a_{IB}^{-1} - a_\pm^{-1}\right) \,.
\]
The detailed derivation is reported in appendix \ref{sec:FrequencyDerivation}.
It contains also a discussion of the case $\lambda^2<0$, where principal value integrals
have to be used.

}

Solving (\ref{eq:lambda}) numerically for different parameters, we
find that it has 
\begin{enumerate}
\item no solution in the attractive regime, so only $\lambda=0$ is possible
here;
\item one solution for positive real $\lambda^{2}$ in the unstable regime;
\item one solution for negative real $\lambda^{2}$ in the repulsive regime.
\end{enumerate}

{These values} give predictions of the exponential
growth rate or frequency, respectively, which are in perfect agreement
with the numerical simulations.

\begin{figure}
  \includegraphics[width=1\columnwidth]{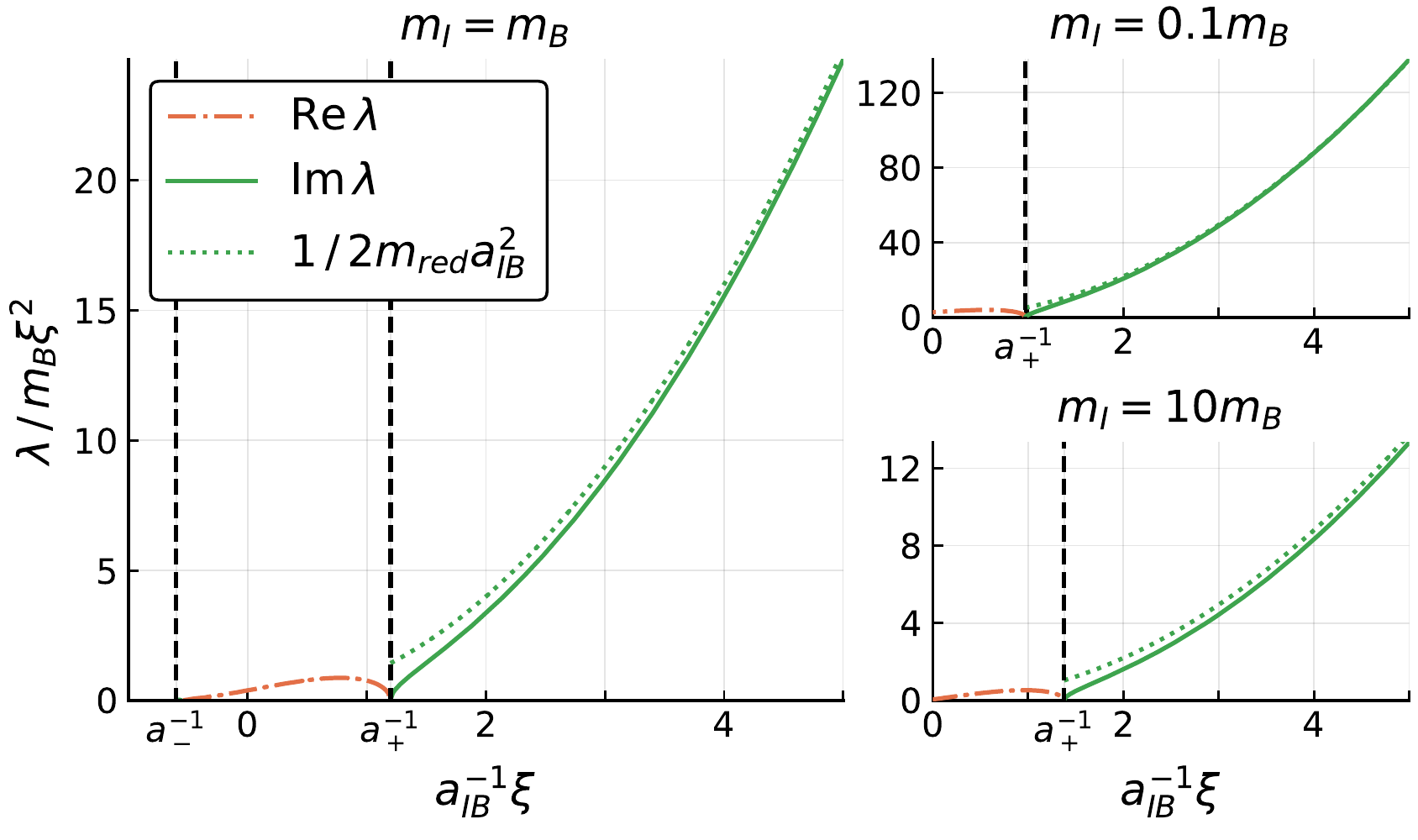}
  \caption{\label{fig:Lambda}Predictions for exponential growth rate
    $\Re\lambda$ { (dash-dotted line)} and oscillation frequency
    $\Im\lambda$ { (solid line)} in the
    long-time limit for the unstable and repulsive regime. The
    oscillation frequency is well approximated by
    $1/2m_{\text{red}}a_{IB}^{2}$, the energy of the two-body bound
    state. As the small figures show, this remains true for different
    mass ratios (note the difference in vertical scale).  Parameters
    are as in Fig.~\ref{fig:Density Profiles} with a hard momentum
    cutoff at $\Lambda=1000\xi^{-1}$.}
\end{figure}

Figure~\ref{fig:Lambda} shows $\lambda$ over a range of different
scattering lengths.
{ In the last section, we have shown
that for an infinitely heavy impurity in a non-interacting BEC, the
oscillation frequency is given by the energy of the two-body bound state, $1/2m_{\text{red}}a_{IB}^{2}$.
We find that this is still a very good approximation for the general case
away from the resonance, i.e. in the repulsive region where our theory applies.
In particular, the frequency depends only on quantities of the impurity-boson scattering problem.
Only close to the resonance, deviations become visible. Here, the
many-body environment leads to a shift of the bound state energy and
consequently, the oscillation frequency starts to depend on properties
of the BEC as well.
}

\begin{figure*}
  \includegraphics[width=0.8\textwidth]{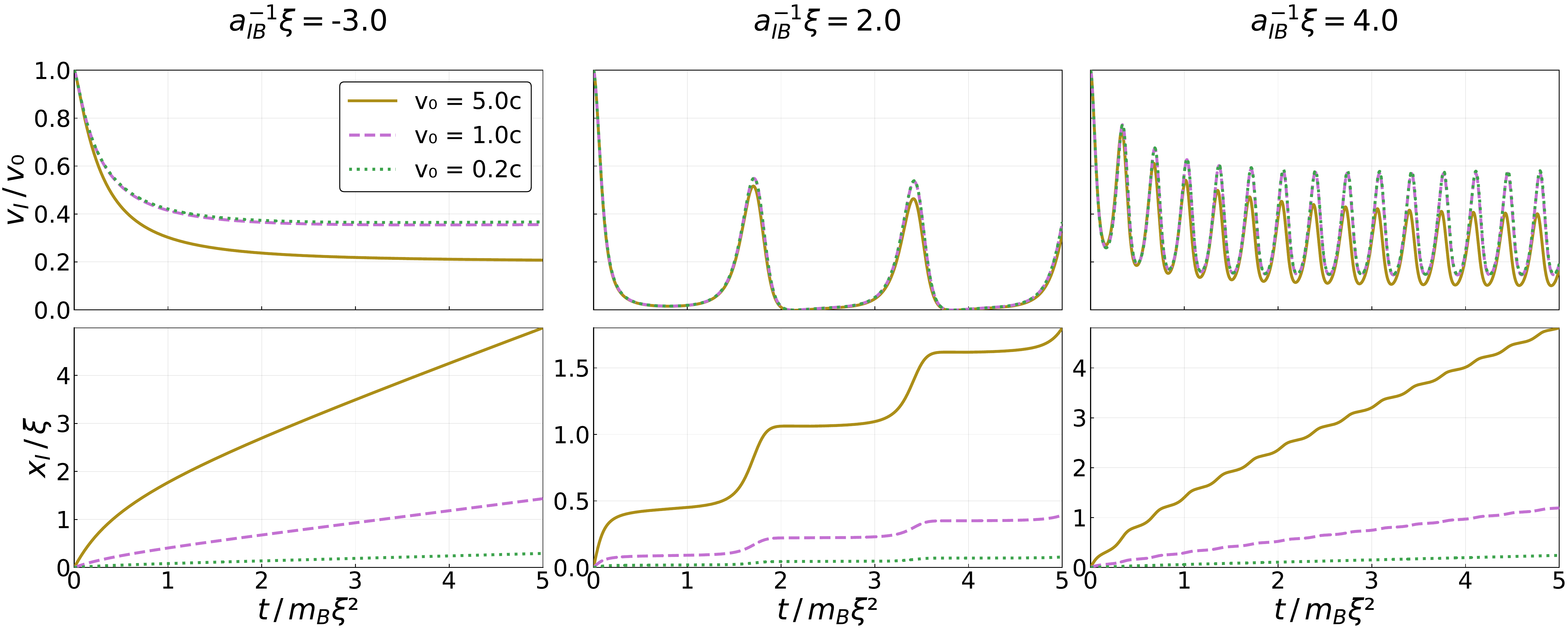}
  \caption{\label{fig:Momentum and Position}Time evolution of velocity
    and position of the impurity for one attractive and two repulsive
    scattering lengths.  In the velocity plots, the curves for
    $v_{0}=1.0c$ and $v_{0}=0.2c$ lie on top of each other. Parameters
    are as in Fig.~\ref{fig:Density Profiles} with a soft cutoff
    $e^{-3k^{2}/\Lambda^{2}}$ and $\Lambda=100\xi^{-1}$.}
\end{figure*}

\begin{figure*}
\includegraphics[width=0.8\textwidth]{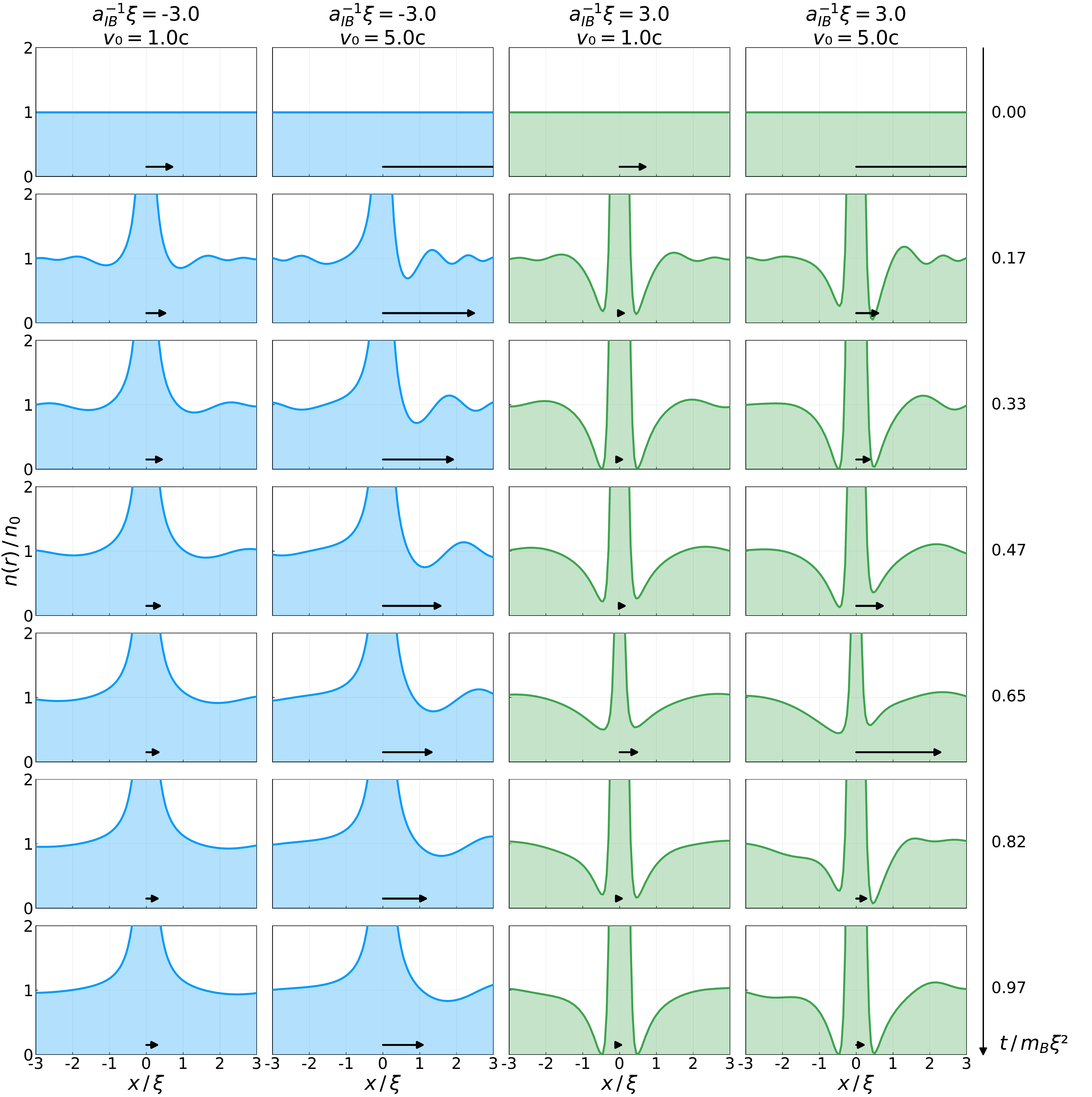}
\caption{\label{fig:density profiles moving}Density profiles for
  nonzero initial momentum along the axis of motion, for attractive
  (left) and repulsive interaction (right). Arrows indicate the
  impurity velocity.  Asymmetry is clearly visible above the speed of
  sound only. Same parameters as in Fig.~\ref{fig:Density Profiles}.}
\end{figure*}

\subsection{Moving Impurity}

We now turn to the case where the impurity has an initial velocity
$\boldsymbol{v_{0}}=\boldsymbol{p_{0}}/m_{I}$. This case has been
investigated for the Fröhlich Hamiltonian in \cite{Shashi2014} using
the coherent-state variational ansatz and in \cite{Grusdt2018} with a time-dependent
renormalization group method. In the latter reference, also the full
Hamiltonian was investigated on the attractive side of the Feshbach
resonance and it was argued that in this case, the second order terms
lead only to a shift of the inverse scattering length. On the repulsive
side, this is not true since the Fröhlich Hamiltonian cannot describe
molecule formation.

In Fig.~\ref{fig:Momentum and Position} we show the time evolution
of the impurity velocity $\boldsymbol{v_{I}}(t)=\boldsymbol{v_{0}}-\frac{1}{m_{I}}\int\frac{d^{3}\boldsymbol{k}}{\k{2\pi}^{3}}\boldsymbol{k}\abs{\alpha_{\boldsymbol{k}}}^{2}$
and position $\boldsymbol{x_{I}}(t)=\int_{0}^{t}\,\boldsymbol{v_{I}}(t')dt'$
(according to Ehrenfest's theorem). In the attractive regime, the
behaviour is simple: If the total momentum is not too high, it converges
to a non-zero final value, which agrees with the stationary solution. {
(Note that a slow-down of the impurity even when its velocity is already
below the speed of sound does not contradict Landau's theory, which makes
predictions for the stationary state. Here, the quench into a far-from
equilibrium state introduces enough interaction energy to excite phonons even
when the impurity is slower than the speed of sound.)}
This matches the picture of a polaron with an increased effective
mass. If the total momentum is too large such that no stationary solution
exists, the velocity converges to the speed of sound. Note, however,
that close to the resonance, it has been predicted that quantum fluctuations
beyond the coherent-state ansatz lead to an enhanced damping or even recoil effects,
cf.~\cite{Grusdt2018}.

On the repulsive side, the behaviour is different: the velocity is
oscillating with the same frequency as the density profile and boson
number. Indeed, the
velocity is smallest when the boson number is largest, which corresponds to a high effective
mass of the impurity.
The effect is most striking close to
the critical scattering length $a_{+}^{-1}$: Here the impurity velocity
quickly reaches zero but has periodic revivals.

On both sides of the resonance, the initial velocity does not matter
much as long as it is below or close to the speed of sound: it leads
only to a rescaling of the velocity at later times. This is also reflected
in the density profiles (Fig.~\ref{fig:density profiles moving}),
which are still symmetric around the impurity. Above the speed of
sound, however, the number of attracted bosons is increased, leading
to a faster decay of the velocity. The density profiles now become
asymmetric with some depletion in front of the impurity.

\section{Discussion}

We investigated the dynamics of polaron formation in a BEC after a
quench, focusing on real space density profiles and the behaviour
for positive scattering lengths. These could not be investigated in
previous works that used the Fröhlich Hamiltonian.

We found that three regions of qualitatively different behaviour exist,
where the strong-coupling region is unstable, as expected from the
stationary analysis in \cite{Grusdt2017}. The fact that the instability
persists even in the limit of heavy impurities is a hint that Bogoliubov
theory is not adequate to investigate the strong-coupling regime:
the deformation of the BEC is too important for the Bogoliubov approximation
to hold. {Our results are thus most reliable away from this critical
region.}

For positive scattering lengths, oscillations can be observed in the
expectation values of many observables and we presented a way to compute
their frequency. {For an infinitely heavy impurity in a non-interacting
BEC, it is exactly given by the energy of the two-body bound state, $1/2m_{\text{red}}a_{IB}^{2}$,
while for the general case, this is still a very good approximation.
In contrast to the case of the Fermi polaron, these oscillations do not
dephase due to coupling to the continuum because they occur in a
coherent bound state. Nevertheless, Bose-Bose interactions beyond Bogoliubov theory
likely lead to damping, but on an independent time scale, that is determined
by properties of the BEC only. In the experimentally relevant case of a weakly
interacting BEC, it will be slow and many oscillations are expected to be observable.
A quantitative estimate is, unfortunately, not possible from within the
theory and requires beyond-Bogoliubov methods.
}

Remarkably, these oscillations are present even in the impurity velocity,
leading to striking ``stop-and-go'' polaron trajectories. The effect
is most pronounced for strong coupling when oscillations are slow
compared to the velocity relaxation: Here the position is advanced
in steps. It will be interesting to see if this can be observed in
experiments.

In position space, the positive scattering length leads to a halo
of reduced condensate density around the impurity, whose size corresponds
to a scattering length and is independent of the mass. This is a version
of a bubble polaron, where the impurity has, however, still a core
of increased density around it and the profile oscillates in intensity.
At certain times, the depletion is even perfect, which was not visible
in ground state calculations {\cite{Ardila2015}}.

Experimentally, the spatial structure of the Bose polaron could be either
detected by direct imaging {on a scale of $a_{IB}$}. Alternatively, both the impurity RF spectra and
Ramsey spectroscopy of the contrast \cite{Mistakidis2018} are
sensitive to {oscillations in }the local density.
{
In the case of a ${}^6$Li impurity in a BEC of $^{133}$Cs atoms \cite{Ulmanis2016}, typical parameters are on the order
of $n_0 a_{BB}^3 = 1.5\cdot 10^{-5}$, $a_{IB} = 100 \si{nm}$ and $a_{BB}=8\si{nm}$. 
The time scale of the oscillations is then $2\pi{2m_\text{red}a_{IB}^2 \over \hbar} \simeq 10 \si{\mu s}$.
On the other hand, the time scale of phonon decay due to beyond-Bogoliubov terms is a
subleading effect of order of $(n_0a_{BB}^3)^{-{1 \over 2}}$ slower than the BEC time scale 
${m_B \xi^2 \over \hbar}$ and therefore of order $10\si{ms}$ for typical parameters.
We therefore expect that a large number of oscillations can be observed.
}

It will be interesting for future work to investigate how the system
behaves for strong coupling where the coherent state ansatz becomes
unstable. However, suitable techniques still have to be developed.
Within Bogoliubov theory, the so-called correlated gaussian wave functions
are a promising way since they should become exact in the limit of
heavy impurities. On the other hand, it will be important to find
out in which region Bogoliubov theory is not reliable any more and
how the system can be described in this region.

\begin{acknowledgments}
We thank Richard Schmidt and Matthias Weidemüller for useful discussions.
This work is part of the DFG Collaborative Research Centre SFB 1225
ISOQUANT.
\end{acknowledgments}

{

\appendix

\section{Undamped oscillations in Bogoliubov Theory\label{sec:OscillationMechanism}}

In the main text, we have shown that an infinitely heavy impurity in a
non-interacting BEC is subject to undamped oscillations due to the presence
of coherently bound states. Here, we argue that this remains true
if Bose-Bose interactions are included within Bogoliubov approximation, i.e., up
to second order in the boson operators with non-zero momentum. Consequently, any
decay of oscillations that one may physically expect can only be due to
beyond-Bogoliubov terms. Since these are proportional to the Bose-Bose coupling
strength $g_{BB}$, the time scale of the decay will be given by properties of the
BEC only. In this appendix we go beyond the rest of the paper in that we do not
restrict the Hilbert space to coherent states but consider the exact dynamics of the
Bogoliubov theory.

The Hamiltonian reads
\begin{align*}
  H_{\text{Bog}} = & \sumprime_{\boldsymbol k, \boldsymbol q}
            \left[ \left({k^2 \over 2m_B} + g_{BB}n_0\right) \delta_{\boldsymbol k, \boldsymbol q} +
              {g_{IB} \over V} \right] a_{\boldsymbol k}^\dagger a_{\boldsymbol q} \\
           + & g_{BB} n_0 \sumprime_{\boldsymbol k} a_{\boldsymbol k}^\dagger a_{- \boldsymbol
              k}^\dagger + a_{\boldsymbol k} a_{- \boldsymbol k} \\
           + & g_{IB} \sqrt{n_0 \over V} \sumprime_{\boldsymbol k} a_{\boldsymbol k}^\dagger + a_{\boldsymbol k}
\end{align*}
after substituting $a_0, a_0^\dagger \rightarrow \sqrt{n_0}$ and dropping third and fourth order
interaction terms but before applying the Bogoliubov transformation.

The first two lines, i.e., the quadratic parts, can be diagonalized by a generalized Bogoliubov
transformation, see \cite{Kain2018}. (In this reference, the transformation is applied on top
of the usual Bogoliubov transformation. We found it simpler to use only one transformation in
total.) This defines new Bogoliubov quasiparticle operators $c$:
\begin{align*}
  c_E &= A_{Ek} a_k + B_{Ek} a_k^\dagger \\
  a_k &= C_{kE} c_E + D_{kE} c_E^\dagger \\
\end{align*}
where 
\begin{align*}
  C_{kE} A_{Eq} + D_{kE} \overline{B_{Eq}} &= \delta_{kq} \\
  C_{kE} B_{Eq} + D_{kE} \overline{A_{Eq}} &= 0 \, ,
\end{align*}
such that
\begin{equation}
  H_{\text{Bog}} = \sum_E E c_E^\dagger c_E + E v_E c_E^\dagger + E \overline{v_E} c_E \, .
  \label{eq:c Hamiltonian}
\end{equation}
The values of $E$ can be thought of as two-body energies that are shifted by the presence
of the many-body environment.
Our analysis is not rigorous in that we do not prove the existence of such a
transformation---while one will be able to find matrices $A$ and $B$ such that (\ref{eq:c Hamiltonian}) holds,
the real question is if the so defined operators $c$ admit a vacuum state $\ket{0}_c$
with $c_E\ket{0}_c = 0$ for all $E$. We expect that this is the case at least outside the unstable
region and that one eigenstate with negative energy exists on the repulsive side, c.f.
\cite{Kain2018}. In fact, above transformation might also be
non-unitary, but in this case, an analogous expression to (\ref{eq:c Hamiltonian}) holds after
transforming to a non-orthogonal basis, which poses no problem.

As for the non-interacting case, the Hamiltonian is of the form
\begin{align*}
  H &= \sum_E d_E^\dagger d_E + const. \\
\end{align*}
for shifted quasi-particles $d_E = c_E + v_E$.
This allows us to compute the time-evolution of the Boson operators $a$ by writing them in terms
of $d$, applying the time evolution operator, and writing the result again in terms of $a$. We
obtain
\begin{align*}
e^{iHt} a_k e^{-iHt} =
  \sum_{E,q} & C_{kE} \left( A_{Eq} a_q + B_{Eq} a_q^\dagger \right) e^{-iEt} \\
       + & D_{kE} \left( \overline{A_{Eq}} a_q^\dagger + \overline{B_{Eq}} a_q \right) e^{iEt} \\
       + & C_{kE} v_E \left( e^{-iEt} - 1 \right) \\
       + & D_{kE} \overline{v_E} \left(e^{iEt} - 1\right) \, .
\end{align*}
Again, the continuum part dephases in the long-time limit, leaving only terms with $E = E_B'$ where
$E_B'$ is the negative energy eigenvalue in (\ref{eq:c Hamiltonian}). As for the non-interacting
case, the time-evolution does not just lead to an oscillating overall phase, but to terms of the
form
$\left( e^{-iEt} - 1 \right)$ that remain visible as interferences in the expectation values of observables.
Moreover, terms with $\exp(-i E t)$ can combine to give oscillations with a frequency of $2E_B'$.
These might, of course, be rather small in amplitude for a weakly interacting
Bose gas. Crucially, the inclusion of Bose-Bose interactions in Bogoliubov approximation has not led
to a damping of the oscillations already present in the non-interacting case, but only to a
frequency shift $E_B \rightarrow E_B'$.

\section{Derivation of the Oscillation Frequencies\label{sec:FrequencyDerivation}}

In this appendix, we derive equation (\ref{eq:lambda}) for the oscillation frequencies
in the repulsive regime and discuss its poles for $\lambda^2 < 0$.
This is done by finding the asymptotic solutions of the
projected Schrödinger equation (\ref{eq:DiffEq}).

As already stated in the main text, we use the ansatz
\begin{align*}
C_{1} & =\sum_{\lambda}A_{\lambda}e^{\lambda t} \\
C_{2} & =\sum_{\lambda}B_{\lambda}e^{\lambda t}
\end{align*}
with finitely many complex $\lambda$, subject to the condition $\Re \lambda \ge 0$.
The prefactors must fulfill
$A_{\overline{\lambda}}=\overline{A_{\lambda}}$ and $B_{\overline{\lambda}}=\overline{B_{\lambda}}$
to ensure that $C_1$ and $C_2$ are real. Inserting into (\ref{eq:DiffEq}) yields
\[
\alpha_{k}(t)=s_{k}e^{-i\Omega_{k}t}+\sum_{\lambda}b_{k\lambda}e^{\lambda t} \, .
\]
The coefficients $\lambda$ are fixed by
\[
b_{k\lambda}=-\frac{W_{k}A_{\lambda}+iW_{k}^{-1}B_{\lambda}}{\Omega_{k}-i\lambda}
\]
while the $s_{k}$ depend on the full history of the system and are therefore not
determined by the asymptotic solution.
}

Re-inserting the expression for $\alpha$ into the definitions of
$C_{1}$ and $C_{2}$ leads to
\begin{align}
\sum_{\lambda}A_{\lambda}e^{\lambda t}= & g_{IB}\sqrt{n_{0}}\nonumber \\
+ & g_{IB}\int\frac{d^{3}\boldsymbol{k}}{(2\pi)^{3}}W_{k}^{\phantom{-1}}\Re\Big(s_{k}e^{-i\Omega_{k}t}+\sum_{\lambda}b_{k\lambda}e^{\lambda t}\Big)\nonumber \\
\sum_{\lambda}B_{\lambda}e^{\lambda t}= & g_{IB}\int\frac{d^{3}\boldsymbol{k}}{(2\pi)^{3}}W_{k}^{-1}\Im\Big(s_{k}e^{-i\Omega_{k}t}+\sum_{\lambda}b_{k\lambda}e^{\lambda t}\Big)\,.\label{eq:ABsum}
\end{align}

These equations should be regarded only as determining the solution
asymptotically because the integrals over $e^{-i\Omega_{k}t}$ will
decay while no finite sum of exponentials $e^{\lambda t}$ with all
$\Re\lambda\ge0$ can ever be decaying.  But since we are interested in
the long-time limit, we can ignore the oscillatory integrals. The need
to drop these terms simply reflects the fact that a system never
exactly reaches its asymptotic solution but only comes arbitrarily
close.

We want to use the linear independence of $e^{\lambda t}$ with different
$\lambda$, but first, the $\Re$ and $\Im$ need to be expanded as
$2\Re\sum_{\lambda}b_{k\lambda}e^{\lambda t}=\sum_{\lambda}(b_{k\lambda}e^{\lambda t}+\overline{b_{k\lambda}}e^{\bar{\lambda}t})=\sum_{\lambda}(b_{k\lambda}+\overline{b_{k\overline{\lambda}}})e^{\lambda t}$
and similarly for $\Im$. We then find
\begin{align*}
\lambda=0: &  & A_{0} & =g_{IB}\left(\sqrt{n_{0}}+\int\frac{d^{3}\boldsymbol{k}}{(2\pi)^{3}}W_{k}\Re b_{k0}\right)\\
 &  & B_{0} & =g_{IB}\int\frac{d^{3}\boldsymbol{k}}{(2\pi)^{3}}W_{k}^{-1}\Im b_{k0}\\
\lambda\ne0: &  & 2A_{\lambda} & =g_{IB}\int\frac{d^{3}\boldsymbol{k}}{(2\pi)^{3}}W_{k}\left(b_{k\lambda}+\overline{b_{k\overline{\lambda}}}\right)\\
 &  & 2iB_{\lambda} & =g_{IB}\int\frac{d^{3}\boldsymbol{k}}{(2\pi)^{3}}W_{k}^{-1}\left(b_{k\lambda}-\overline{b_{k\overline{\lambda}}}\right)\,.
\end{align*}
Recall that $A_{\lambda}$ and $B_{\lambda}$ need not be real even
though the sums in (\ref{eq:ABsum}) are. Also note that these equations
would not be true for $\Re\lambda<0$ because the oscillating integrals
could not be ignored.

Using the expressions for $b_{k\lambda}$ and the relations 
\begin{align*}
g_{IB}^{-1}+\int\frac{d^{3}\boldsymbol{k}}{(2\pi)^{3}}\frac{W_{k}^{\pm2}}{\Omega_{k}} & =\frac{\mu}{2\pi}\left(a_{IB}^{-1}-a_{\pm}^{-1}\right)\\
 & =:\Delta_{\pm}
\end{align*}
we arrive at
\begin{widetext}
\begin{align*}
\lambda=0: &  & A_{0} & =\frac{\sqrt{n_{0}}}{\Delta_{+}}\\
 &  & B_{0} & =0\\
\lambda\ne0: &  & A_{\lambda}\left(\Delta_{+}-\lambda^{2}\int\frac{d^{3}\boldsymbol{k}}{(2\pi)^{3}}\frac{W_{k}^{2}}{\Omega_{k}(\Omega_{k}^{2}+\lambda^{2})}\right) & =\lambda B_{\lambda}\int\frac{d^{3}\boldsymbol{k}}{(2\pi)^{3}}\frac{1}{\Omega_{k}^{2}+\lambda^{2}}\\
 &  & B_{\lambda}\left(\Delta_{-}-\lambda^{2}\int\frac{d^{3}\boldsymbol{k}}{(2\pi)^{3}}\frac{W_{k}^{-2}}{\Omega_{k}(\Omega_{k}^{2}+\lambda^{2})}\right) & =-\lambda A_{\lambda}\int\frac{d^{3}\boldsymbol{k}}{(2\pi)^{3}}\frac{1}{\Omega_{k}^{2}+\lambda^{2}}\,.
\end{align*}
Written in this way, all integrals are UV convergent. Multiplying
the last two equations finally leads to equation (\ref{eq:lambda}) for $\lambda^{2}$,
independent of $A_{\lambda}$ and $B_{\lambda}$: 
\begin{equation}
\left(\Delta_{+}-\lambda^{2}\int\frac{d^{3}\boldsymbol{k}}{(2\pi)^{3}}\frac{W_{k}^{2}}{\Omega_{k}(\Omega_{k}^{2}+\lambda^{2})}\right)
\left(\Delta_{-}-\lambda^{2}\int\frac{d^{3}\boldsymbol{k}}{(2\pi)^{3}}\frac{W_{k}^{-2}}{\Omega_{k}(\Omega_{k}^{2}+\lambda^{2})}\right) 
=-\lambda^{2}\left(\int\frac{d^{3}\boldsymbol{k}}{(2\pi)^{3}}\frac{1}{\Omega_{k}^{2}+\lambda^{2}}\right)^{2}
\, . \label{eq:lambdaAppendix}
\end{equation}
\end{widetext}

\subsection*{The case of $\lambda^2 < 0$}

In the above expressions, many of the integrals do in fact not exist
if $\lambda$ is purely imaginary since the integrands have a pole
at $k=k_{c}$ in this case. What does exist, however, are the Cauchy
principal value (PV) integrals 
\[
\mathcal{P}\int_{0}^{\Lambda}\ldots=\lim_{\epsilon\rightarrow0}\k{\int_{0}^{k_{c}-\epsilon}\ldots+\int_{k_{c}+\epsilon}^{\Lambda}\ldots}\,.
\]
Such integrals are not invariant under coordinate transformations
because the way in which the pole is approached is crucial, so it
is not immediately clear how to make sense of (\ref{eq:lambdaAppendix}) for
negative $\lambda^{2}$. This becomes clearer if, instead of making
an ansatz for an asymptotic solution, one considers the time evolution
operator in (\ref{eq:time evolution matrix}). In fact, the product
$H^{(2)}H^{(1)}$ determines the dynamics completely, so it is sufficient
to compute its spectrum. One obtains, once again, equation (\ref{eq:lambdaAppendix}),
where now $\lambda^{2}$ are the eigenvalues of $H^{(2)}H^{(1)}$.
But in the case $\lambda^{2}<0$, one finds that (\ref{eq:lambdaAppendix})
must hold with the integrals replaced by PV integrals in any choice
of coordinates, as long as the same is used in all three integrals.
Therefore, a coordinate transformation will change the value of the
individual integrals, but when it is applied to all of them, the equation
must stay true.

{
As stated in the main text, one finds no solution in the attractive regime
and one solution in the unstable and repulsive regime.
Using PV integrals in a particular choice of coordinates, one may
find a second solution in the two latter cases, but they
are not valid because they change when different coordinates are used.
The valid solution is therefore unique and leads to figure \ref{fig:Lambda}.
}

\bibliographystyle{apsrev_titles_modified}
\bibliography{references}

\end{document}